\documentclass[12pt]{iopart}
\usepackage{graphicx}
\usepackage{iopams}  
\begin{document}

\title{Viscosity and Thermalization}

\author{Derek A. Teaney}

\address{Department of Physics, Brookhaven National Laboratory, Upton, NY 11973-5000}

\ead{dteaney@quark.phy.bnl.gov}

\begin{abstract}
I solve a Relativistic Navier Stokes Model assuming boost 
invariance and rotational symmetry. 
I compare the resulting numerical solutions for 
two limiting models of the shear viscosity. 
In the first model the shear viscosity is made proportional
to the temperature. Thus,  $\eta \propto T/\sigma_0$ where $\sigma_0$ is
some fixed cross section (perhaps $\sigma_0 \sim \Lambda_{QCD}^{-2}$) . 
This viscosity model is typical of the classical Boltzmann simulations
of Gyulassy and Molnar. In the second model the shear viscosity is 
made proportional to $T^3$. This model is typical of 
high temperature QCD. 
When the initial mean free path of the $T^3$ model is four times 
larger than the $T/\sigma_0$ model, the two models of viscosity 
produce the same radial flow. 
This result can be understood with simple scaling arguments.
Thus, the large transport opacity needed in classical Boltzmann simulations
is in part an artifact of the fixed scale $\sigma_0$ in these models. 
\end{abstract}




%
\noindent {\bf 1. Motivation: } The observation of elliptic flow is 
one of the most striking results the heavy ion program  
\cite{v2}. In mid-peripheral collisions the azimuthal anisotropy 
of the produced particles 
$v_2(p_T) \equiv \left\langle \cos(2 \phi) \right \rangle_{p_T} $ rises linearly
as a function of transverse momentum up to $p_T \sim 1.5\,\mbox{GeV}$  
and then flattens and maintains a constant value of  approximately $15\%$.
Ideal hydrodynamics provides an economical description of the observed 
elliptic flow for transverse momentum below $1.5\,\mbox{GeV}$ \cite{hirano}.
However, ideal hydrodynamics  assumes that the transport mean free path  $\ell_{mfp}$ path is so small that  viscous terms can be neglected. 
Simple 
estimates indicate that that in heavy ion 
collisions this assumption  is marginal. It  is hoped that $\ell_{mfp}/L$ is
of order $\sim ~1/5$ and that some semblance of an equilibrium QGP plasma
will be formed. For $\ell_{mfp}/L \sim 1/5$  viscous hydrodynamics should 
provide a semi-quantitative guide to the evolution of the system. 
Viscous hydrodynamics should then explain the systematics of the elliptic
flow measurements. 

Elliptic flow was studied in the fully non-equilibrium framework  of
classical kinetic theory by Gyulassy and Molnar \cite{GM}. 
These authors
varied the cross section between classical particles and computed the 
resulting elliptic flow. The results are surprising.
First, the full kinetic theory reproduces the full shape of the 
$v_2(p_T)$ curve. In particular the kinetic theory describes the linear 
rise with transverse momentum (as expected from hydrodynamics) and 
the subsequent flattening of $v_2(p_T)$ .
The first viscous correction to the thermal distribution function   
clarifies this transition from the hydrodynamic to the kinetic 
regime \cite{Teaneyvis}. In spite of this success, these classical Boltzmann simulations 
suggest that
hydrodynamics {\it is not} responsible for the observed $v_2$. The 
cross sections needed to reproduce the observed elliptic flow are 
$\approx 10\,\mbox{mb}$. With such cross sections the initial momentum 
degradation length is approximately four times smaller than 
the thermal wavelength. This seems impossible.  

However, several differences between between QCD and these 
classical cascade models should be noted. 
First, the classical cascade preserves particle number. 
Second,
the classical simulation have a fixed scale $\sigma_0$.  This
is not so unreasonable -- perhaps $\sigma_0$ is $\Lambda_{QCD}^2$.
To understand the implications  of this fixed scale  let us estimate
the temporal  dependence of the viscosity in different situations.


In the perturbative quark gluon plasma the shear viscosity is proportional $T^3$ as given  by dimension.  Thus the ratio of the shear viscosity to the entropy $\eta/s$ is
a constant up to logarithms. 
In a classical massless gas with  constant cross sections the shear viscosity 
is given by  $\eta = 1.264\, T/\sigma_0$ Thus the temperature dependence of 
the shear viscosity  in the classical model \cite{GM}  differs from high temperature QCD.  
Simple scaling arguments indicate that a model  with a constant cross 
section is unlikely to thermalize.
For a Bjorken expansion, the condition for 
hydrodynamics to be valid is: 
$\Gamma_s/\tau \ll 1$ where 
where $\Gamma_s$ is the sound attenuation length $\frac{4}{3} \eta/(e+p)$  .

Next, I estimate how $\Gamma_s/\tau$ evolves as a function of time  
during a Bjorken expansion
for the two limiting  models of viscosity discussed above.  For a Bjorken
expansion $T\propto 1/\tau^{1/3}$ and $n \propto 1/\tau$.
Then, when the viscosity is proportional to $T^3$ we  
find  that the system comes closer to equilibrium as a function of time
\begin{eqnarray*}
 \frac{\Gamma_s}{\tau} \sim \frac{1}{\tau T}  \sim \frac{1}{\tau^{2/3}} 
&\;\;\;\;& \mbox{(1D Bjorken with $\eta\propto T^3$)} . 
\end{eqnarray*}
In contrast for a classical gas with conserved particle number 
$e + p = 4\, n T$ and constant cross section the degree of thermalization 
remains constant as a function of time
\begin{eqnarray*}
   \frac{\Gamma_s}{\tau} \sim \frac{1} { \tau  n \sigma_0} 
   \sim \mbox{Const} &\;\;\;\;& \mbox{ (1D Bjorken $\eta\propto T/\sigma_0$) } .
\end{eqnarray*}
Similar instructive arguments may be given when the system expands in three dimensions.  
The conclusion remains the same. When the viscosity is proportional to $T^3$
the system is much more likely to thermalize than when other scales such as 
$\sigma_0$ and $\Lambda_{QCD}$ enter the problem.


\noindent {\bf 2. Viscous Solutions. } It has been understood for some time that the
relativistic Navier stokes equations can not be solved directly. However,
this problem is cosmetic rather than fundamental in nature \cite{Lindblom}. 
A myriad of different hydrodynamic models \cite{HCO,Models} 
can be solved numerically with 
considerable effort \cite{DT}.  These models all give the same solution
up to corrections which are proportional to $\left(\Gamma_s/\tau \right)^2$. 
The stress energy tensor is always close to its canonical form 
$T^{ij} \sim \eta\, \left(\partial^{i} v^{j} + \partial^{j} v^{i} - \frac{2}{3}\delta^{ij}\, \partial_l v^{l}\right) $. 
These model equations have no greater validity than the Navier Stokes
equation. The hydrodynamic model used here was 
inspired by  a model with appealing mathematical structure 
due to Ottinger \cite{HCO}. 

\begin{figure}
\begin{center}
\includegraphics[height=3.0in, width=3.0in]{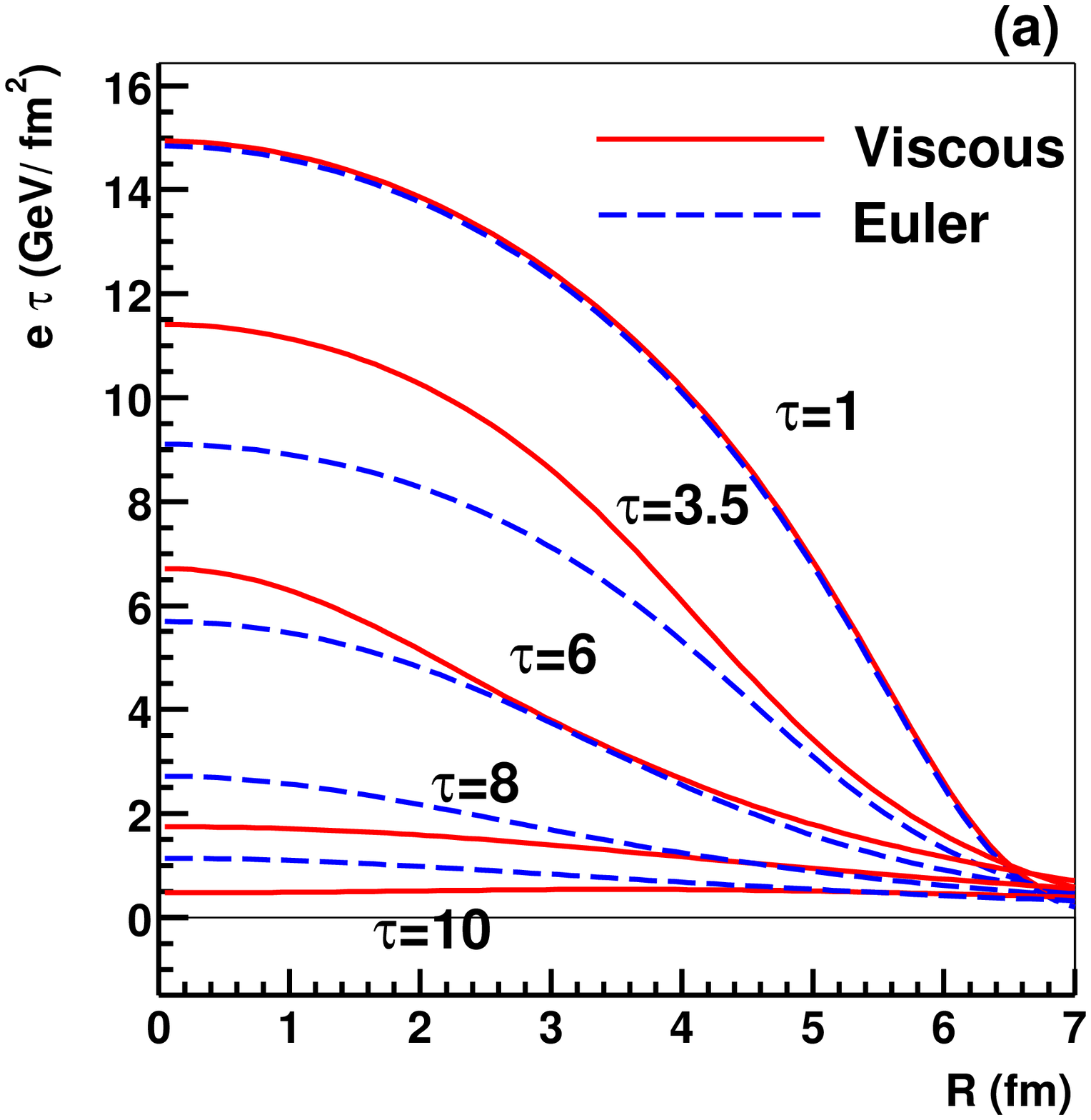}
\includegraphics[height=3.0in, width=3.0in]{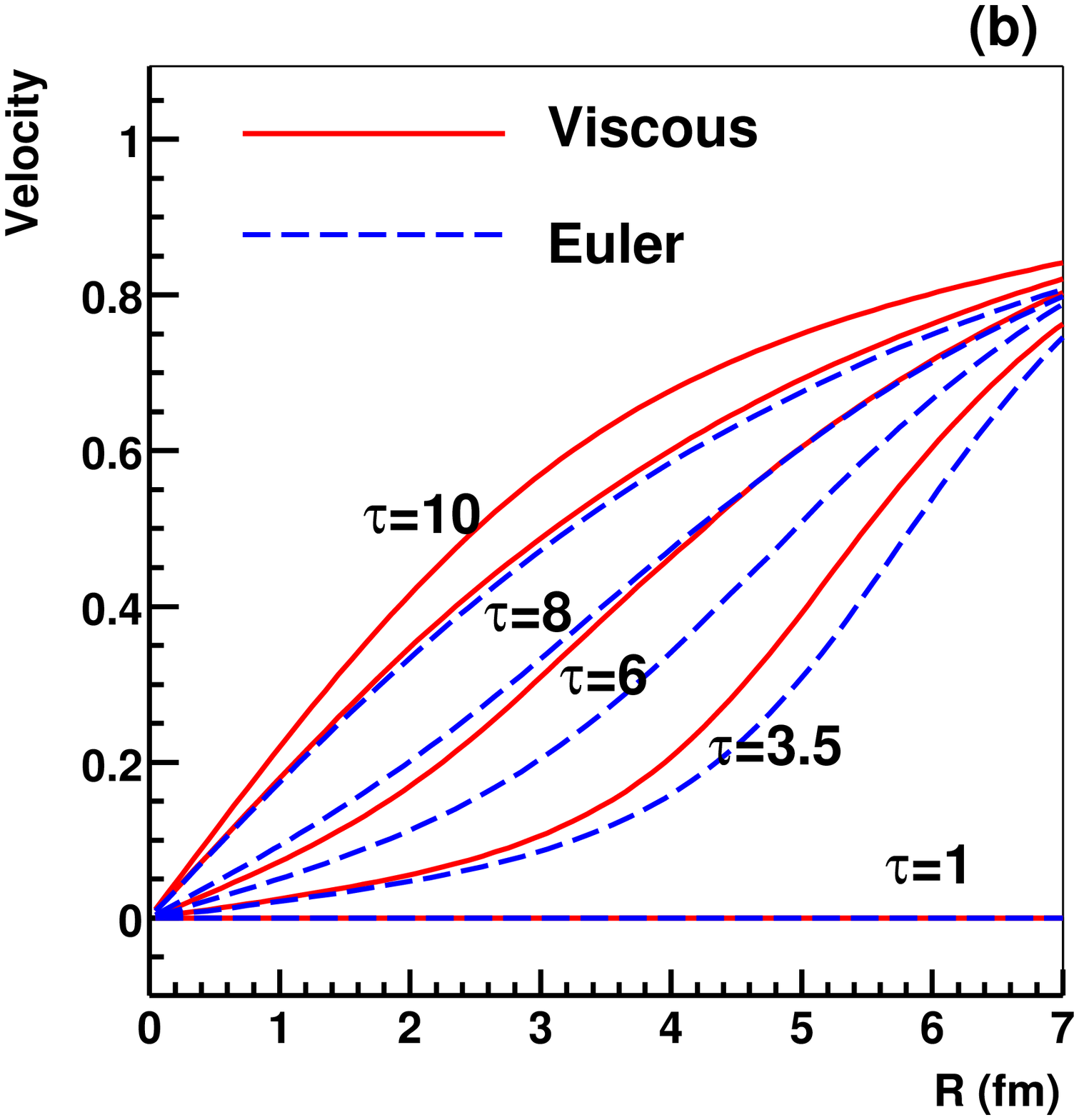}
\caption{A comparison of the viscous and inviscid (Euler) solutions 
for a Bjorken expansion with radial symmetry. The initial conditions
are from P. Kolb's hydrodynamic model \cite{Kolb}. The 
equation of state is $p = \frac{1}{3} e$. The viscosity is proportional 
to the entropy $\eta/s = \frac{1}{5} $. (a) The energy density multiplied by $\tau$
for various times (b) The fluid velocity
for various times. }
\label{solution}
\end{center}
\end{figure}

First, I solve a Navier Stokes Model (the Ottinger Model \cite{HCO}) assuming boost 
invariance and radial symmetry. The initial conditions are 
taken from P. Kolb's inviscid hydrodynamic calculations \cite{Kolb} 
which reproduces the observed spectra and multiplicity. 
The viscosity is made proportional to the entropy\, $\eta/s = 1/5$.  
The viscous solution 
is compared with the inviscid (Euler) solution shown in Fig.~\ref{solution}. 
Examine the energy density in Fig.~\ref{solution}(a). 
First the viscous solution does less longitudinal 
work since the longitudinal pressure is reduced by the 
longitudinal expansion (see e.g. \cite{Teaneyvis}). 
Consequently, the energy density initially 
decreases more slowly for the viscous case. 
However,  the transverse pressure is increased by longitudinal expansion.
 This causes the transverse flow  
to rise more rapidly in the viscous case as seen in Fig.~\ref{solution}(b).
This larger transverse flow velocity subsequently causes
the energy density to fall more rapidly in the viscous case. By 
a time of $\approx 6\,\mbox{fm}$ the viscous and inviscid solutions  
are similar. In summary, viscous corrections do not integrate to yield an
order one change to the inviscid flow.

Next, I compare two simple models for the viscosity.  
The first model for the viscosity is taken from a classical ideal
gas with a constant cross sections. In this case
$\eta = 1.264 \frac{T}{\sigma_0}$ with $\sigma_0=10\,\mbox{mb}$. This 
model of the viscosity has been studied within the domain of kinetic
theory by Gyulassy and Molnar \cite{GM}.  

The second model for the viscosity is referred to as the Minimal 
Model below. In this model we take
\begin{equation}
 \eta = \left\{ \begin{array}{lc}  
 \rm 1.264\, \frac{T}{\sigma_0} &  \mbox{ for }  e < e_c      \\
  \rm     \frac{1}{5} \, s           &  \mbox{ for }  e > e_c  \end{array}  \right.
\end{equation}
where 
$e_{c} = 1\,\mbox{GeV/fm}^3$ and $\sigma_{0}=10\,\mbox{mb}$. This 
model of the shear viscosity has $\eta\propto T^3$ for high temperatures
but has a fixed scale $\sigma_0$ 
(i.e. $\Lambda_{QCD}^{-2}$) at low temperatures.  The shear 
viscosity of the Minimal Model  is always larger than the fixed 
cross section model. 

\begin{figure}
\begin{center}
\includegraphics[height=3.0in, width=3.0in]{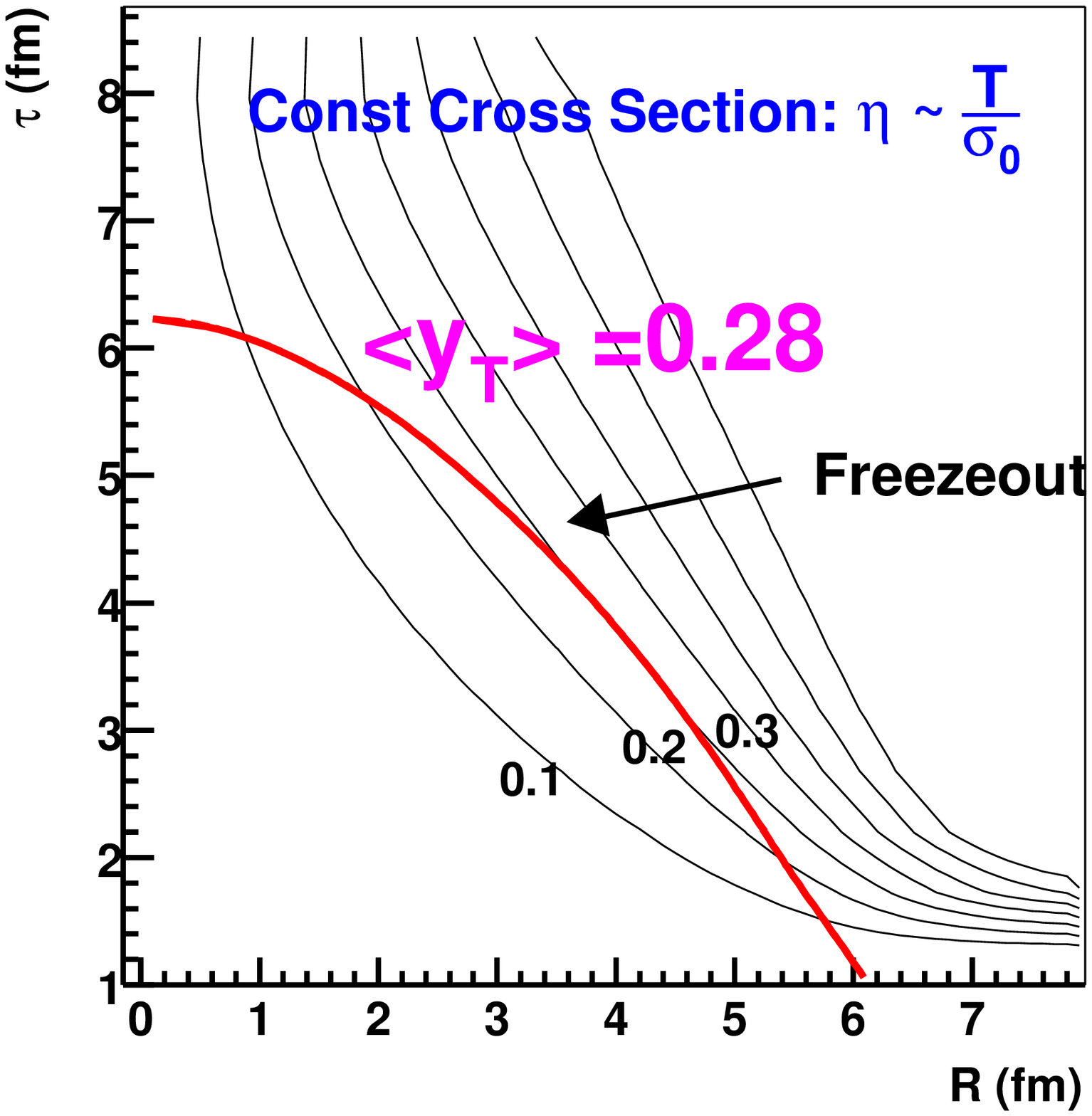}
\includegraphics[height=3.0in, width=3.0in]{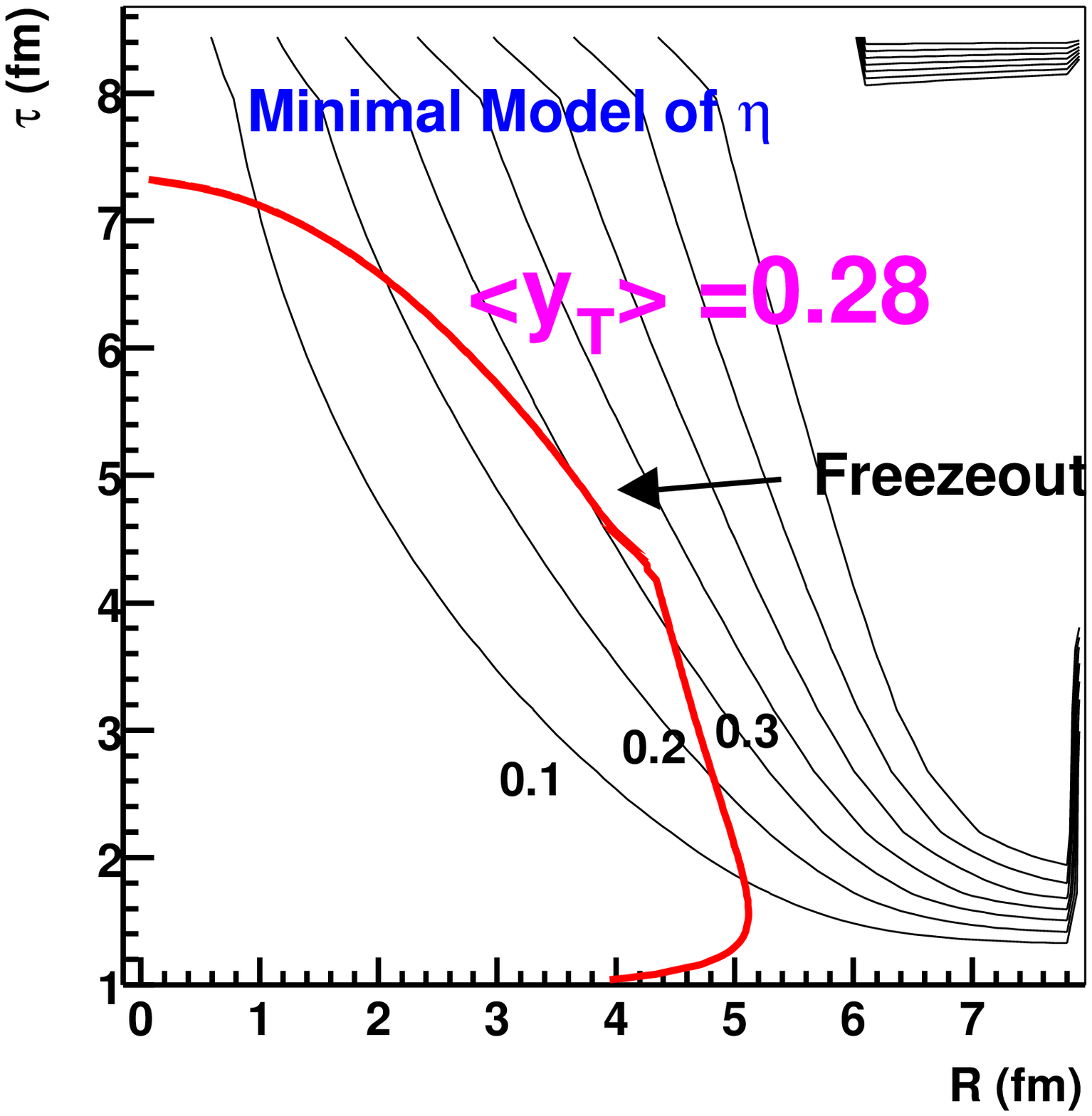}
\caption{A comparison of two models of the shear viscosity. In the 
first model the shear viscosity $\eta =1.264 T/\sigma_0$. 
The minimal model of viscosity is described in the text and  has $\eta\propto
T^3$ when the energy density exceeds $1.0\,\mbox{GeV/fm}^3$. 
In both case $p = \frac{1}{3}e$. The thin 
lines are lines of constant transverse rapidity with values $0.1,\, 0.2,\, 0.3\,...$ The thick line labeled Freezeout is where the viscous ``correction''
is half of the pressure. $\left\langle y_T \right\rangle$ denotes the entropy 
weighted mean transverse rapidity along the freezeout curve. }
\label{models}
\end{center}
\end{figure}

The viscous hydrodynamic solutions to these models are illustrated
in Fig. ~\ref{models}. The two models of viscosity give approximately 
the same solution although the viscosity in the Minimal Model is initially four
times larger than the fixed cross section model.
Eventually, the viscosity becomes large and the system freezes out. 
The curve labeled ''freezeout'' indicates when the viscous correction
becomes equal to half of the hydrodynamic pressure. The mean transverse
relativistic velocity $\left\langle y_T \right\rangle$ is calculated along the 
''freezeout'' curve using the Cooper-Frye formula and weighting each
surface element by the entropy. 
The two models of the shear viscosity give approximately the same  
transverse flow as can be seen be comparing $\left\langle y_T \right\rangle$ in
each case.   Thus the large transport opacity 
needed in classical Boltzmann simulations is in part an artifact of the fixed
scale $\sigma_0$ introduced into the problem.
\vspace{0.5cm}

\end{document}